\documentclass[aps,prl,showpacs,twocolumn,superscriptaddress]{revtex4}
\usepackage{amsmath}
\usepackage{amssymb}
\usepackage{epsfig}
\usepackage{color}
\usepackage[colorlinks]{hyperref}
\hypersetup{colorlinks,citecolor=red,linkcolor=blue,urlcolor=blue}
\usepackage{graphicx,amsmath}
\begin{document}
\title{Universal $T/B$ scaling behavior of heavy fermion compounds}
\author{V. R. Shaginyan}\email{vrshag@thd.pnpi.spb.ru}
\affiliation{Petersburg Nuclear Physics Institute, NRC Kurchatov
Institute, Gatchina, 188300, Russia}\affiliation{Clark Atlanta
University, Atlanta, GA 30314, USA}\author{A. Z. Msezane}
\affiliation{Clark Atlanta University, Atlanta, GA 30314, USA}
\author{J.~W. Clark} \affiliation{McDonnell Center
for the Space Sciences \& Department of Physics, Washington
University, St.~Louis, MO 63130, USA} \affiliation{Centro de
Investiga\c{c}\~{a}o em Matem\'{a}tica e Aplica\c{c}\~{o}es,
University of Madeira, 9020-105 Funchal, Madeira, Portugal}
\author{G. S.  Japaridze}\affiliation{Clark Atlanta University, Atlanta,
GA 30314, USA}
\author{Y. S. Leevik} \affiliation{National Research University
Higher School of Economics, St.Petersburg, 194100, Russia}

\begin{abstract}
In our mini-review, we address manifestations of $T/B$ scaling
behavior of heavy-fermion (HF) compounds, where $T$ and $B$ are
respectively temperature and magnetic field.  Using experimental
data and the fermion condensation theory, we show that this scaling
behavior is typical of HF compounds including HF metals,
quasicrystals, and quantum spin liquids. We demonstrate that such
scaling behavior holds down to the lowest temperature and field
values, so that $T/B$ varies in a wide range, provided the HF
compound is located near the topological fermion condensation
quantum phase transition (FCQPT). Due to the topological properties
of FCQPT, the effective mass $M^*$ exhibits a universal behavior,
and diverges as $T$ goes to zero. Such a behavior of $M^*$ has
important technological applications. We also explain how to
extract the universal scaling behavior from experimental data
collected on different heavy-fermion compounds. As an example, we
consider the HF metal $\rm YbCo_2Ge_4$, and show that its scaling
behavior is violated at low temperatures. Our results obtained show
good agreement with experimental facts.
\end{abstract}

\maketitle

\section{Introduction}

Topological approach is a powerful method to gain information about
a wide class of physical systems. Knowledge of the topological
properties allows us to improve a general knowledge about physical
systems without solving specific equations, which describe concrete
systems and are often very complicated. As usually, the microscopic
approach to a heavy fermion (HF) metal (for example, computer
simulations) gives only particular information about specific
solids, but not about universal features, inherent in the wide
class of HF compounds. HF compounds can be viewed as the new state
of matter, since their behavior near the topological fermion
condensation quantum phase transition (FCQPT) acquire important
similarities, making them universal. The idea of this phase
transition, forming experimentally discovered flat bands, started
long ago, in 1990 \cite{khodel:1990,vol,graph}. At first, this idea
seemed to be a curious mathematical exercise, and now it is proved
to be rapidly expanding field with uncountable applications
\cite{khodel:1990,graph,vol,khodel:1994,volov_gr,shaginyan:2010,bamusia:2015,shaginyan:2016,book2020}.

The scaling behavior of HF compounds is a challenging problem of
condensed matter physics
\cite{shaginyan:2010,mats:2011,coleman:2015,gegen:2016,coleman:2019}.
It is generally assumed that scaling with respect to $T/B$
(temperature-magnetic field ratio) is related to a quantum critical
point (QCP) that represents the endpoint of a phase transition
being tuned to $T=0$ by such control parameters as magnetic field,
pressure, and composition of the heavy-fermion compounds. As soon
as the tuned endpoint of the phase transition reaches $T=0$, it
becomes a quantum phase transition (QPT). At QCP involved quantum
fluctuations like valence, magnetism, etc can take place and
influence on the properties of system in question
\cite{coleman:2015,gegen:2016}. Fluctuations can also occur at
second-order phase transitions, but in all cases the temperature
range of these fluctuations is very narrow \cite{lanl}; in
contrast, $T/B$ scaling can span a few orders of magnitude in $T/B$
\cite{shaginyan:2010,bamusia:2015,shaginyan:2016}. An attendant
problem to be addressed by theory stems from the experimental
finding that scaling behavior can take place without both QCP
realization and effective mass $M^*$ divergence \cite{gegen:2016}.
The divergence of effective mass $M^*$ at $T\to 0$ is of crucial
importance for understanding technological applications of quantum
materials. For example, the divergence leads to the high heat
capacity $C$ of quantum material, while under the application of
magnetic field both $M^*$ and $C$ diminishes. As a result, one can
exploit this property constructing low temperature refrigerators.
To solve these problems, one needs to have a reliable theoretical
framework for analysis of experimental facts related to the scaling
behavior.

A universal $T/B$ scaling behavior is generated by quasiparticles
belonging to flat bands. These flat bands can be preformed by van
Hove singularities and finally formed by inter-particle interaction
generating topological FCQPT \cite{prb:2013}. In narrow electronic
bands in which the Coulomb interaction energy becomes comparable to
the bandwidth, interactions drive the topological FCQPT; as a
result, at $T=0$ flat bands are emerged, see e.g.
\cite{shaginyan:2010,bamusia:2015,shaginyan:2016,prb:2013}. Such
flat bands in twisted graphene have been experimentally observed
see e.g. \cite{graph}. Thus, we can safely use the model of
homogeneous HF liquid, since we consider a behavior controlled by
flat bands and related to the scaling of quantities such as the
effective mass, heat capacity, magnetization, etc. As a result, the
scaling properties are defined by momentum transfers that are small
compared to momenta of the order of the reciprocal lattice length.
The high momentum contributions can therefore be ignored by
substituting the lattice for the jelly model; this observation is
in a good agreement with experimental facts collected on many HF
metals \cite{shaginyan:2010,bamusia:2015,book2020}. In our case
quasiparticles are well defined excitations see e.g.
\cite{shaginyan:2010,bamusia:2015,khod:2010}, and the divergence of
the effective mass $M^*$ is not related to $Z\to 0$ (as it is can
be assumed, see e.g. \cite{varma}), where $Z$ is the quasiparticle
amplitude. The divergence is defined by both the emergence of an
extended Van Hove singularities, pre-forming flat bands, and the
Coulomb interaction, giving rise to strong correlations; as a
result, at $T=0$ the electronic dispersion becomes flat at the
chemical potential $\mu$, topologically transforming the Fermi
surface into Fermi volume, see e.g.
\cite{shaginyan:2010,bamusia:2015,book2020,khodel:1990,shag1998,kats:14,kats,prb:2013}.

In our mini-review, we show that the fermion condensation (FC)
theory, which entails the topological FCQPT, provides the
appropriate framework for describing and analyzing the universal
scaling behavior of HF compounds
\cite{shaginyan:2010,bamusia:2015,book2020,khodel:1990,vol,khodel:1994,volov_gr}.
We predict that $T/B$ scaling behavior can be observed in a wide
range of $T/B$ values, provided the given HF compound is located
near a topological FCQPT. Violation of $T/B$ scaling at the lowest
values of $T/B$ is a signal that the given HF compound is situated
before the topological FCQPT on the $T-B$ phase diagram, and hence
exhibits Landau Fermi-Liquid (LFL) behavior at sufficiently low
temperatures. We consider the HF metal $\rm YbCo_2Ge_4$, and show
that its scaling behavior is violated at low temperatures. The
results of the FC theory are in good agreement with experimental
observations collected on different strongly correlated Fermi
systems like HF metals, quasicrystals and quantum magnets, holding
quantum spin liquids \cite{shaginyan:2010,bamusia:2015,book2020}.
As a result, the FC theory is useful tool when projecting
technological applications of quantum materials representing by HF
compounds.

\section{Scaling behavior of the effective mass
near the topological FCQPT}

One of the main experimental manifestations of the topological
FCQPT phenomenon is the scaling behavior of the physical properties
of HF compounds located near such a phase transition. To understand
this scaling behavior on a sound theoretical basis, we begin with a
brief description of the associated behavior exhibited by the
effective mass $M^*$ in the framework of a { homogeneous} HF liquid
\cite{shaginyan:2010}. This simplification avoids the complications
associated with the anisotropy of solids and focuses of both the
thermodynamic properties and the non-Fermi-liquid (NFL) behavior by
calculating the effective mass $M^*(T,B)$ as a function of
temperature $T$ and magnetic field $B$
\cite{shaginyan:2010,bamusia:2015,shaginyan:2016}, based on the
Landau formula for the quasiparticle effective mass $M^*(T,H)$. The
only modification introduced is that the effective mass is no
longer approximately constant but now depends on temperature,
magnetic field, and other parameters such as pressure, etc. {We
note that the FC theory is a good established theory based on the
density functional theory; in that case, the Landau functional
$E[n({\bf p})]$ and the corresponding Eq. \eqref{HC1_15} are also
derived in the same frameworks,  therefore, being exact, see e.g.
\cite{shaginyan:2010,bamusia:2015,book2020,shag1998,khodel:1994}.
Here $n({\bf p})$ is the quasiparticle distribution function.}

At finite temperatures and magnetic fields, Landau's equation takes
the form
\cite{shaginyan:2010,shaginyan:2016,khodel:2011,landau:1956}
\begin{equation}\label{HC1_15}
\frac{1}{M^*_{\sigma}(T, H)}=\frac{1}{M}
\end{equation}
$$+\sum_{\sigma_1}\int\frac{{\bf p}_F{\bf p}}{p_F^3}F_
{\sigma,\sigma_1}({\bf p_F},{\bf p}) \frac{\partial n_{\sigma_1}
({\bf p},T,H)}{\partial{p}}\frac{d{\bf p}}{(2\pi)^3}
$$
in terms of $n_\sigma({\bf p})$ and the quasiparticle interaction
$F_{\sigma_1,\sigma_2}$. Here $M$ can be represented by the bare
electron mass or by the enhanced bare like mass at van Hove
singularity. The single-particle spectrum $\varepsilon({\bf p},T)$
is a variational derivative
\begin{equation}
\varepsilon_{\sigma}({\bf p})=\frac{\delta E[n({\bf p})]}{\delta
n_{\sigma}({\bf p})},\label{EN_15}
\end{equation}
of the system energy $E[n_{\sigma}({\bf p})]$ with respect to the
quasiparticle distribution (or occupation numbers) $n_\sigma({\bf
p}$ for spin $\sigma$, which in turn is related to the spectrum
$\varepsilon_\sigma$({\bf p}) by
\begin{equation}
n_{\sigma}({\bf p},T)=\left\{ 1+\exp\left[\frac{(\varepsilon({\bf
p},T)-\mu_{\sigma})}T\right]\right\} ^{-1}.\label{HC25}
\end{equation}
In our case, the chemical potential $\mu$ depends on spin due to
the Zeeman splitting, $\mu_{\sigma}=\mu\pm \mu_BB$, where $\mu_B$
is the Bohr magneton. The magnetic field $B$ appears in
Eq.~\eqref{HC25} via the ratio $\mu_{\sigma}/T=(\mu\pm \mu_BB)/T$.
We note that \eqref{HC1_15} and \eqref{EN_15} are exact equations
\cite{bamusia:2015,shag1998}. In our case, the Landau interaction
$F$ is fixed by the condition that the system is situated at a
FCQPT. Its sole purpose is to bring the system to the topological
FCQPT point, where $M^*\to\infty$ at $T=0$ and $B=0$, altering the
topology of the Fermi surface, transforming it to a volume such
that the effective mass acquires temperature and field dependence
\cite{shaginyan:2010,khodel:2011,clark:2005}.  Provided the Landau
interaction is an analytic function, at the Fermi surface the
momentum-dependent part of the Landau interaction can be
parameterized as a truncated power series $F=aq^2+ bq^3+cq^4+...$,
where ${\bf q}={\bf p}_1-{\bf p}_2$, the variables ${\bf p}_1$ and
${\bf p}_2$ are momenta, and $a,b$, and $c$ are fitting parameters
defined by the condition that the system is at a FCQPT point.

\begin{figure}
\begin{center}
%\vspace*{-2.5cm}
\includegraphics [width=0.47\textwidth]{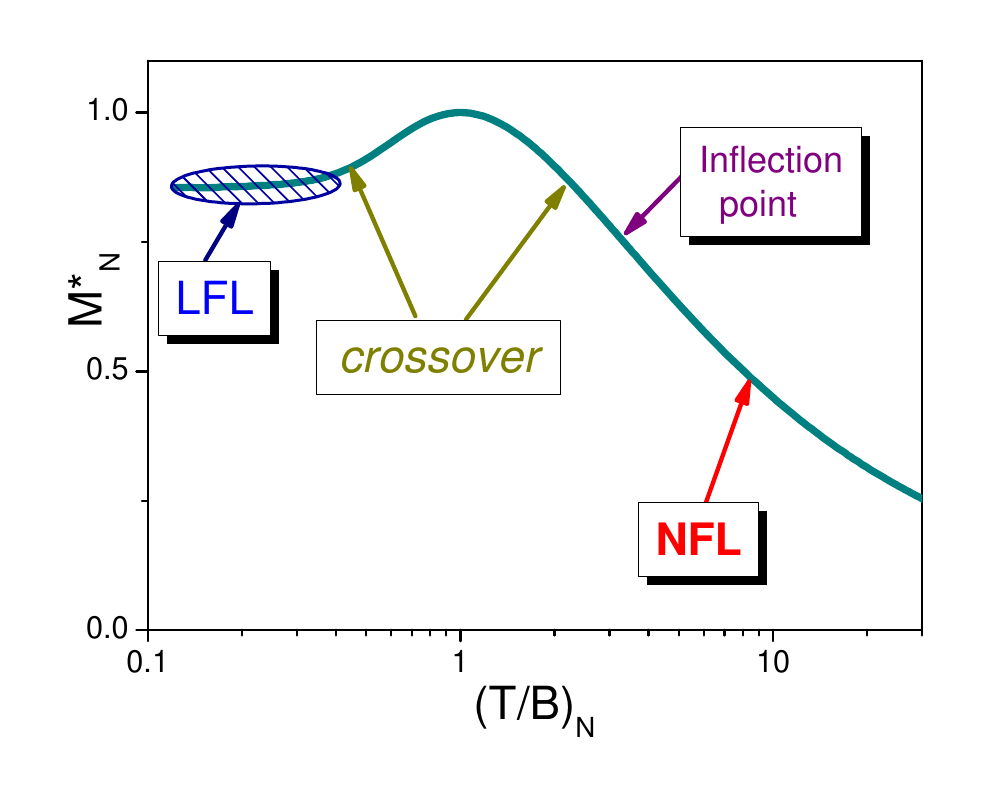}
%\vspace*{-0.75cm}
\end{center}
\caption{(Color online). Scaling behavior of the dimensionless
effective mass $M_N^*$ versus dimensionless variable $(T/B)_N$.
Scaling of thermodynamic properties is defined by $M^*_N$; see
Eq.~\eqref{kuka1}. $M^*_N$ is a function of $T_N\propto(T/B)_N\sim
(T/B)/(T/B)_M$, as it follows from Eq.~\eqref{TMB7}. Solid curve
depicts the scaling behavior $M^*_N$ versus normalized temperature
$T_N$ as a function of magnetic field, given by Eqs.~\eqref{UN27}
and \eqref{UN215}. Clearly, at finite $T_N<1$ the normal Fermi
liquid regime is realized. At $T_N\sim 1$ the system enters a
crossover state, and at growing temperatures exhibits NFL behavior.
The LFL, crossover, inflection point, and NFL behavior are
indicated by the arrows.  The LFL behavior is additionally shown by
the hatched area. }\label{fig9}
\end{figure}

\begin{figure}[!ht]
\begin{center}
\includegraphics [width=0.47\textwidth]{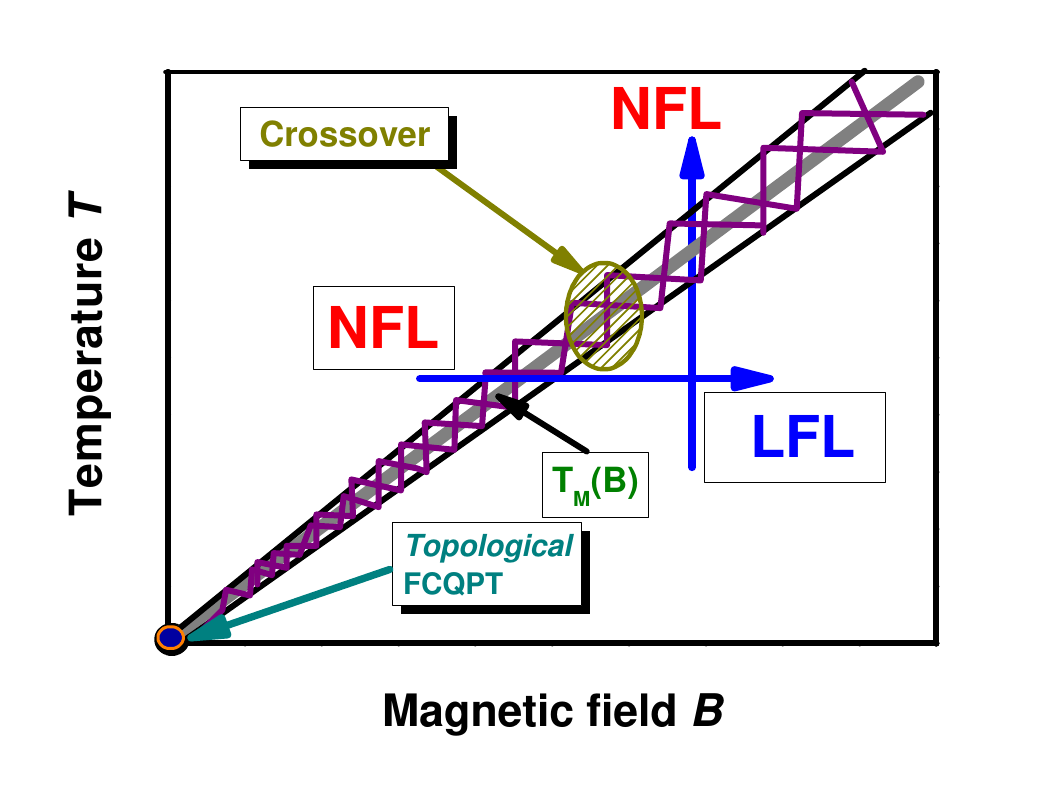}
%\vspace*{-0.75cm}
\end{center}
\caption{(Color online). Schematic $T-B$ phase diagram of a HF
compound, with magnetic field $B$ as control parameter.  The
hatched area corresponds to the crossover domain at $T_M(B)$, given
by Eq.~\eqref{TMB7}. At fixed magnetic field and elevated
temperature (vertical arrow) there is a LFL-NFL crossover. The
horizontal arrow indicates a NFL-LFL transition at fixed
temperature and elevated magnetic field. The topological FCQPT
(shown in the panel) occurs at $T=0$ and $B=0$, where $M^*$
diverges.}\label{fig10}
\end{figure}

Direct inspection of Eq.~\eqref{HC1_15} shows that at $T=0$ and
$B=0$, the sum of the first and second terms on the right side
vanishes, since $1/M^*(T\to0)$ goes to zero when the system is
located at the FCQPT point.  Given a Landau interaction analytic
with respect to momenta variables, at finite $T$ the right side of
Eq.~\eqref{HC1_15} is proportional $F^{\prime}(M^*)^2T^2$, where
$F^{\prime}$ is the first derivative of $F(q)$ with respect to $q$
at $q\to0$.  Results for the corresponding integrals can be found
in textbooks -- see especially Ref.~\cite{lifshitzem:2002}.  At any
rate, we have $1/M^*\propto (M^*)^2T^2$ and arrive at
\cite{shaginyan:2010,bamusia:2015}
\begin{equation}
M^*(T)\simeq a_TT^{-2/3}.\label{MT15}
\end{equation}
At finite temperatures, application of a magnetic field $\mu_B
B\gg k_BT$ drives the system to the LFL regime with
\begin{equation}
M^*(B)\simeq a_BB^{-2/3},\label{MB15}
\end{equation}
where $a_T$ and $a_B$ are parameters and $k_B$ the Boltzmann
constant. It follows from Eq. \eqref{MB15} that heat capacity
$C\propto M^*(B)T$ of quantum materials strongly depends on $B$,
this property can be used when projecting e.g. low temperatures
refrigerators.

If the system is still located before the FCQPT, the effective mass
is finite $M^*=M_0$, and Eq.~\eqref{MB15} must be adjusted, since
at $B\to 0$ the effective mass $M^*$ does not diverge; thus
\begin{equation}
M^*(B)\simeq a_B(B_0+B)^{-2/3}.\label{MB15F}
\end{equation}
From Eq.~\eqref{MB15F} it follows that at $B\to0$ the effective
mass becomes $M_0$. Therefore, when $B\gg B_0$, the effective mass
$M^*$ depends on the magnetic field in accordance with
Eq.~\eqref{MB15}, since the contribution coming from $B$ defines
the behavior of $M^*$. As a result, at $B\gg B_0$ we can replace
Eq.~\eqref{MB15} by the equivalent equation
\begin{equation}
M^*(B)\simeq M_0+a_BB^{-2/3}.\label{MBC15}
\end{equation}
In the case of the HF liquid, these observations allow for
construction of an approximate solution of Eq.~\eqref{HC1_15} in
the form $M^*=M^*(B,T)$ that satisfies both Eqs.~\eqref{MT15} and
\eqref{MB15}. Introduction of ``internal'' scales simplifies the
problem under consideration, allowing us to eliminate the
microscopic structure of the HF compounds under consideration
\cite{shaginyan:2010,bamusia:2015}. To establish such ``internal''
scales, we observe that near the FCQPT, the effective mass
$M^*(B,T)$ reaches a maximum $M^*_M$ at a certain temperature
$T_{M}\propto B$.  (See later comments on Eqs.~\eqref{HC25} and
\eqref{MB15} and Fig.~\ref{fig9}).  To conveniently measure the
effective mass and temperature versus magnetic field $B$, we
introduce the scales $M^*_M$ and $T_{M}$, generating new variables
$M^*_N=M^*/M^*_M$ (normalized effective mass) and $T_N=T/T_{M}$
(normalized temperature). In the vicinity of FCQPT, the normalized
effective mass $M^*_N(T_N)$ is well approximated by a universal
function \cite{shaginyan:2010,bamusia:2015}
\begin{equation}M^*_N(T_N)\approx c_0\frac{1+c_1T_N^2}{1+c_2T_N^{8/3}}.
\label{UN27}
\end{equation}
Here, $T_N=T/T_{M}\propto T/B$ (see Fig.~\ref{fig9}) and
$c_0=(1+c_2)/(1+c_1)$, with $c_1$ and $c_2$ free parameters. We
stress that values of $M^*_M$ and $T_{M}$ are defined by the
microscopic structure of the HF compound under study, while the
normalized values $M^*_N$ and $T_N$ demonstrate the universal
scaling exhibited by HF compounds located near the topological
FCQPT, since this scaling is determined by the nature of both the
phase transition and the model of homogeneous HF liquid; we note
that these observations are in good agreement with experimental
facts collected on HF compounds, see e.g.
\cite{shaginyan:2010,bamusia:2015,shaginyan:2016,book2020}. From
Eqs.~\eqref{MB15} and \eqref{UN27} it follows that
\begin{equation}
M^*_M\propto B^{-2/3}\propto T_M^{-2/3};\, T/B\simeq
T/T_N.\label{MAX}
\end{equation}
The Landau interaction $F_{\sigma,\sigma_1}(q)$ appearing in
Eq.~\eqref{HC1_15} can produce the characteristic topological form
of the spectrum $\varepsilon(p)-\mu\propto(p-p_b)^2(p-p_F)$, with
$(p_b<p_F)$ and $(p_F-p_b)/p_F\ll1$, leading to $M^*\propto
T^{-1/2}$ and creating a quantum critical point
\cite{khodel:2005:A}. The same critical point is generated by the
interaction $F(q)$ as represented by a non-analytic but
integrable-over-$x$ function with $q=\sqrt{p_1^2+p_2^2-2xp_1p_2}$
and $F(q\to0)\to \infty$
\cite{shaginyan:2010,khodel:1994,shaginyan:2016}. Both cases lead
to $M^*\propto T^{-1/2}$, and Eq.~\eqref{MT15} becomes
\begin{equation}
M^*(T)\simeq a_T T^{-1/2}.\label{MTT_15}
\end{equation}
In the same way, we obtain
\begin{equation}
M^*(B)\simeq a_B B^{-1/2},\label{MBB_15}
\end{equation}
in terms of parameters $a_T$ and $a_B$.

Taking into account the fact that Eq.~\eqref{MTT_15} leads to a
spiky density of states (DOS), with the spiky character fading away
under increasing temperature as observed in quasicrystals
\cite{shaginyan:2013,widmer:2009,deguchi:2012}, we note that the
general form of $\varepsilon(p)$ produces the behavior of $M^*$
given by Eqs.~\eqref{MTT_15} and \eqref{MBB_15}. This is realized
in quasicrystals, which can be viewed as a generalized form of
common crystals \cite{shaginyan:2013}. We note further that the
behavior $1/M^*\propto\chi^{-1}\propto T^{1/2}$ is in good
agreement with the behavior $\chi^{-1}\propto T^{0.51}$ observed
experimentally in quasicrystals \cite{deguchi:2012,shaginyan:2013}.
Our result $1/M^*\propto T^{1/2}$ is consistent with the robustness
of the exponent $0.51$ under hydrostatic pressure
\cite{deguchi:2012}. This robustness is guaranteed by the unique
singular density of states associated with the topological FCQPT,
which survives under application of pressure
\cite{fujiwara:1991,widmer:2009,fujiwara:1993,deguchi:2012,shaginyan:2013}.
To develop the consequences of the the solution of
Eq.~\eqref{HC1_15} at finite $B$ and $T$ near the FCQPT, we
construct an approximate solution by interpolating between the LFL
behavior described by Eq.~\eqref{MBB_15} and the NFL behavior
described by Eq.~\eqref{MTT_15}, that models the universal scaling
behavior $M^*_N(T_N\propto T/B)$ \cite{shaginyan:2013}
\begin{equation}M^*_N(T_N)
\approx c_0\frac{1+c_1T_N^2}{1+c_2T_N^{5/2}},
\label{UN215}
\end{equation}
with $c_0=(1+c_2)/(1+c_1)$ and $c_1$, $c_2$ as fitting parameters.
Taking into account Eqs.~\eqref{MBB_15} and \eqref{UN215}, we
arrive at
\begin{equation}
M^*_M\propto B^{-1/2}\propto T_M^{-1/2}.\label{MAXX}
\end{equation}

It follows from Eqs.~\eqref{HC25}, \eqref{UN27}, and \eqref{UN215}
that
\begin{equation}
\label{TMB7} T_M\propto B; T_N=\frac{T}{T_M}=
\frac{T}{a_1\mu_BB}\propto \frac{T}{B}\sim
\left(\frac{T}{B}\right)_N.
\end{equation}
Here $a_1$ is a dimensionless factor, $\mu_B$ is the Bohr magneton,
$({T}/{B})_N=({T}/{B})/({T}/{B})_M$, where $({T}/{B})_M$ is the
point at which $M^*_N$ reaches it maximum value $M^*_N=1$, as
illustrated in Fig.~\ref{fig9}. Expression \eqref{TMB7} shows that
Eqs.~\eqref{UN27} and \eqref{UN215} determine the effective-mass
scaling in terms of $T/B$ as well. We conclude from
Eq.~\eqref{TMB7} that since $T_M\propto B$, the curves
$M^*_{N}(T,B)$ merge into a single curve $M^*_{N}(T_N=T/B)$, with
$T_N=T/T_M\propto T/B$, demonstrating the widespread scaling in HF
metals (for example, see \cite{shaginyan:2010,bamusia:2015}). Such
behavior is depicted in Fig.~\ref{fig9}.  We note that
Eqs.~\eqref{UN27} and \eqref{TMB7} allow us to describe the
behavior of the strongly correlated quantum spin liquid (SCQSL)
existing in different frustrated magnets
\cite{shaginyan:2010,bamusia:2015}.

Another important feature of the FC state is that apart from the
fact that the Landau quasiparticle effectively acquires strong
dependence on external factors such as the temperature and magnetic
field, all the fundamental relations inherent in the LFL approach
remain formally intact.  In particular, the famous LFL relation
\cite{landau:1956,lifshitzem:2002,shaginyan:2010,bamusia:2015},
\begin{equation}\label{kuka1}
 M^*(B,T)  \propto \chi(B,T) \propto \frac{C(B,T)}{T},
\end{equation}
still holds. That is, expression \eqref{kuka1} is also valid in the
case of HF compounds located near a topological FCQPT, where the
specific heat $C$, magnetic susceptibility $\chi$, and effective
mass $M^*$ depend on $T$ and $B$. Based on Eq.~\eqref{kuka1}, we
find that the normalized values of $C/T$ and $\chi$ are of the form
\cite{shaginyan:2010,bamusia:2015}
\begin{equation}
\label{TMBN} M^*_N(B,T)=\chi_N(B,T)=\left(\frac
{C(B,T)}{T}\right)_N.
\end{equation}
Thus, Eq. \eqref{TMBN} allows us to reveal the universal scaling
behavior of different HF compounds located near the topological
FCQPT like HF metals, frustrated insulators with SCQSL,
quasicrystals, 2D liquids, etc., see e.g.
\cite{shaginyan:2010,book2020} It is also seen from
Eq.~\eqref{TMBN} that the aforementioned thermodynamic properties
have the same scaling behavior as depicted in Fig.~\ref{fig9}.
Moreover, we shall see below that the thermodynamic properties of
HF metals, SCQSL of frustrated magnets, and the other HF compounds
exhibit the same typical behavior. Based on Eq.~\eqref{UN27} and
Fig.~\ref{fig9}, we can construct the general schematic $T-B$ phase
diagram of SCQSL, reported in Fig.~\ref{fig10}. We assume here that
at $T=0$ and $B=0$ the system is approximately located at a FCQPT
point. At fixed temperature the system is driven by the magnetic
field $B$ along the horizontal arrow (from the NFL to the LFL parts
of the phase diagram).  At fixed $B$ and elevated $T$ the system
moves from the LFL to the NFL regime along the vertical arrow. The
hatched area indicating the crossover between LFL and NFL phases
separates the NFL state from the paramagnetic slightly polarized
LFL state. The crossover temperature $T_M(B)$ is given by
Eq.~\eqref{TMB7}.

\section{T/B scaling in heavy fermion compounds}\label{5_TB}

\begin{figure}[!ht]
\begin{center}
%\vspace*{-0.75cm}
\includegraphics [width=0.47\textwidth]{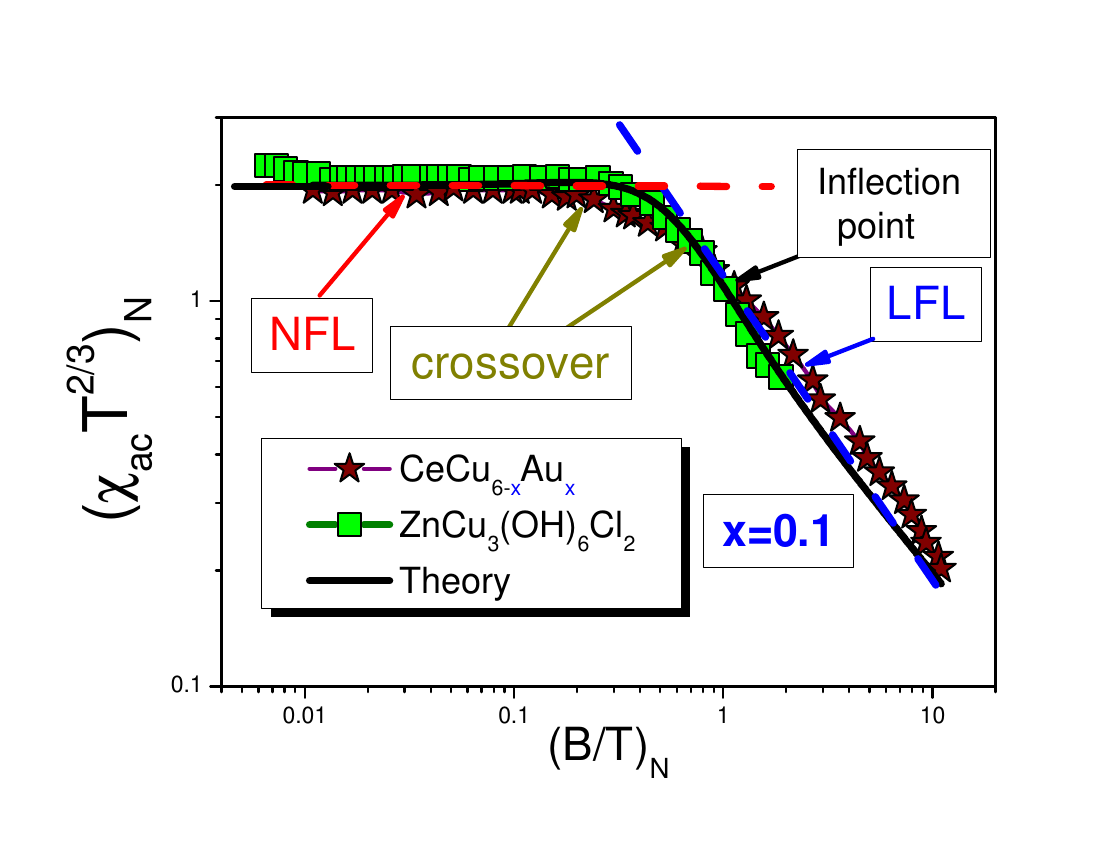}
%\vspace*{-0.75cm}
\end{center}
\caption{(Color online). Universal $B/T$ scaling of strongly
correlated Fermi systems. {Scaling behavior of the HF metal $\rm
CeCu_{6-x}Au_x$ with $x=0.1$ is extracted from data (measured at
different field values $B=0.05, 0.1, 0.3, 0.6, 0.9, 1.05$ T) in
Ref.~\cite{Coleman:2000}, and that of $\rm ZnCu_3(OH)_6Cl_2$
(measured at different field values $B=0.5, 1.0, 3.0, 5.0, 7.0,
10.0, 14.0$ T), from data in Ref.~\cite{helton:2010}.} At $B/T\ll
1$ the systems demonstrate NFL behavior with $\chi\propto M^*$ as
given by Eq.~\eqref{MT15}, i.e., $T^{2/3}\chi\propto$ {\rm const}.
At $B/T\gg 1$ the systems demonstrate LFL behavior with $\chi$ as
given by Eq.~\eqref{MB15}, a decreasing function of $B/T$ (see
Eq.~\eqref{UN215}). The LFL, crossover, inflection point, and NFL
behavior are indicated by the arrows. The broken lines indicate the
asymptotic dependencies in the limits of small $(B/T)_N$ (the NFL
behavior) and large $(B/T)_N$ (the LFL behavior). The theoretical
prediction is represented by solid curve.} \label{fig_05_03_sc}
\end{figure}
The experimentally based scaling behavior of $M^*_N$ so derived is
displayed in Fig.~\ref{fig9}.  Explanation of this scaling,
$M^*_N(T_N)\propto C(B,T)/T$, presents a serious challenge to
theories of the HF compounds.  Most of the current theories analyze
only the critical exponents that characterize $M^*_N(T_N)$ at
$T_N\gg 1$ and thus consider only a part of the problem, missing
the LFL and the transition regime
\cite{shaginyan:2010,bamusia:2015,coleman:2019}. This scaling
behavior of the effective mass $M^*_N$ of HF compounds (or strongly
correlated Fermi systems) is described by Eqs.~\eqref{UN27} and
\eqref{UN215}. It follows then from Eqs.~\eqref{kuka1} and
\eqref{TMBN} that their experimentally observed thermodynamic
properties express the universal scaling behavior revealed by our
analysis.

\begin{figure}[!ht]
\begin{center}
%\vspace*{-0.75cm}
\includegraphics [width=0.47\textwidth]{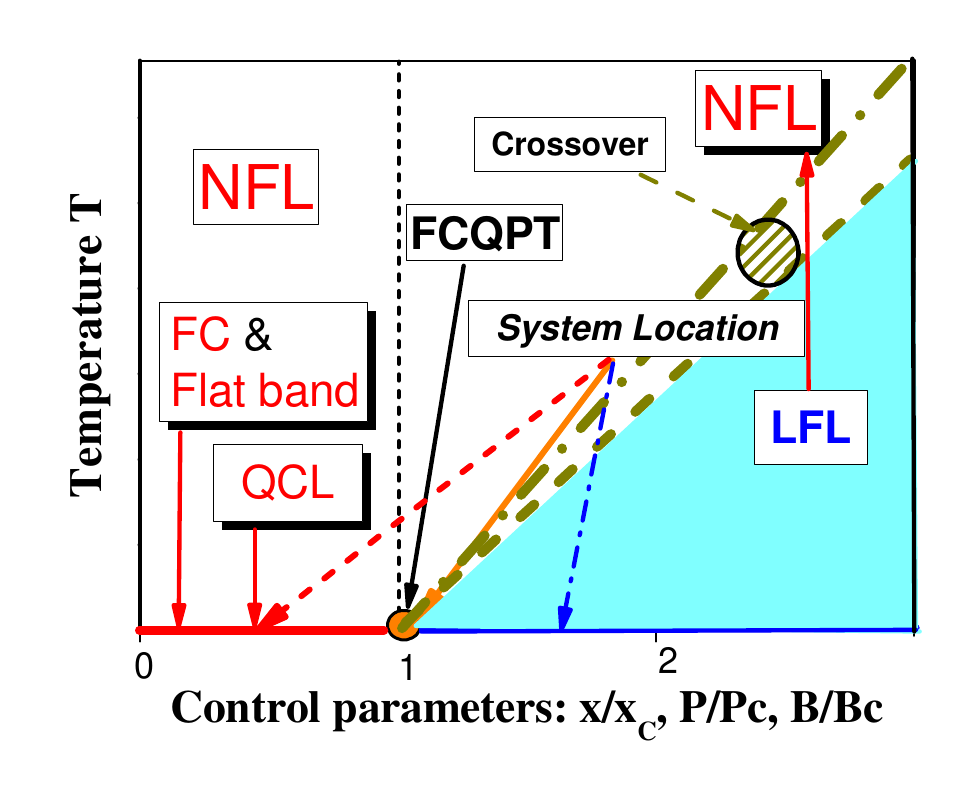}
%\vspace*{-0.75cm}
\end{center}
\caption{(Color online). Schematic diagram of temperature versus
these dimensionless control parameters: normalized pressure
$P/P_c$, composition $x/x_c$, magnetic field $B/B_c$. We assume
that $B_c>0$, if $B_c=0$, see the phase diagram \ref{fig10}. The
solid black line indicates the topological FCQPT point (orange
circle). At $T=0$ and beyond the quantum critical point (to the
left of the orange circle), the system is on the quantum critical
line (QCL) implicating a flat band, as indicated by the red-dashed
arrow. At any finite temperature $T<T_f$ and at elevated $P/P_c>1$,
$x/x_c>1$, $B/B_c>1$, the system enters the crossover and, then,
the LFL region. The blue dash-dot arrow points to the system as
situated before the topological FCQPT, where at $T\to0$ it exhibits
LFL behavior with effective mass $M^*=M_0$.  At elevated magnetic
fields, the behavior of the effective mass is given by
Eq.~\eqref{MBC15}, and the scaling behavior is restored at
$B>B_0$.}\label{fig3}
\end{figure}
In the present context, the HF compounds are taken to represent
strongly correlated Fermi sytems as realized in HF metals,
high-$T_c$ superconductors, quasicrystals, SCQSL of frustrated
magnets and two-dimensional liquids like $^3$He. One can expect
that HF compounds with their extremely diverse composition and
microscopic structure would demonstrate very different
thermodynamic, transport, and relaxation properties. To reveal the
universal scaling behavior of HF compounds, irrespective of
specific properties of individual compounds, we have introduced
internal scales to measure the corresponding thermodynamic
properties, as is done when we consider the scaling behavior of the
effective mass $M^*$. This uniform behavior arises from the fact
that HF compounds are located near a topological FCQPT, generating
their uniform scaling behavior with respect to the effective mass
$M^*$ \cite{shaginyan:2010,shag2017,book2020} (see
Fig.~\ref{fig9}). As an example, Fig.~\ref{fig_05_03_sc} displays
the universal $T/B$ scaling behavior of the HF metal $\rm
CeCu_{6-x}Au_x$ and SCQSL of the frustrated insulator
herbertsmithite $\rm ZnCu_3(OH)_6Cl_2$
\cite{Coleman:2000,helton:2010}. The existence of such universal
behavior, exhibited by various and very distinctive strongly
correlated Fermi systems, supports the conclusion that HF compounds
represent a new state of matter \cite{shag2017,book2020}. In
contrast to the situation for an ordinary quantum phase transition,
this scaling, induced by the topological FCQPT, occurs up to high
characteristic temperature $T_f$, $T<T_f \sim 100$ K, since the NFL
behavior is defined by quasiparticles (with $M_N^*$ given by
Eqs.~\eqref{UN27} and \eqref{UN215}), rather than by fluctuations,
or by Kondo lattice effects
\cite{shaginyan:2010,bamusia:2015,book2020}.

Some remarks are in order here. A strongly correlated Fermi system
can be situated after the topological FCQPT, i.e., on the ordered
side defined by the quantum critical line (QCL), as shown in the
schematic phase diagram \ref{fig3}. As it is shown in Fig.
\ref{fig3}, FCQPT can be tuned by dimensionless control parameters:
Normalized pressure $P/P_c$, composition $x/x_c$, magnetic field
$B/B_c$. Here we assume that the critical magnetic field $B_c>0$.
In the case of $B_c=0$, see Fig. \ref{fig10}, demonstrating that at
$T=0$ and $B=0$ the system is located at the topological FCQPT.
Note that there can be two critical magnetic fields $B_{c1}$ and
$B_{c2}$, as it is in case of the HF metal $\rm Sr_3Ru_2O_7$
\cite{shag:2013,prb:2013}.

As it is seen from Fig. \ref{fig3}, at $T=0$ the crossover region
is absent, and the FC state is separated from the LFL region by the
first order phase transition \cite{shaginyan:2010}, for the FC
state is characterized by special quantum topological number, being
a new type of Fermi liquid \cite{vol}. At $T>0$ there is the
crossover rather than a phase transition \cite{shaginyan:2010}. One
may expect that the $T/B$ scaling is caused by features not related
to the presence of QCP and the divergence of $M^*$ (see e.g.
\cite{gegen:2016}). On the other hand, if the system in question is
located before FCQPT, as indicated by the dash-dot arrow in
Fig.~\ref{fig3}, it exhibits LFL behavior even in the absence of a
magnetic field $B$ at low $T\to 0$. At elevated magnetic fields
reaching $B\gg B_{0}$, Eqs.~\eqref{MB15} and \eqref{MBB_15} are
valid and the scaling behavior returns to that given by
Eqs.~\eqref{UN27} and \eqref{UN215}. Thus, to witness the presence
of { both} the scaling behavior { and} divergence of the effective
mass in measurements on HF compounds, one has to carry out
measurements at sufficiently low temperatures and magnetic fields.
For instance, the HF metal $\rm CeRu_2Si_2$ exhibits NFL behavior
at low temperatures (down to 170 mK) and small magnetic fields
($B\simeq 0.02$ mT) comparable with the magnetic field of the Earth
\cite{takahashi:2003}. Measurements carried out under application
of magnetic fields have led to the incorrect statement that $\rm
CeRu_2Si_2$ demonstrates LFL behavior at low temperatures (see
\cite{takahashi:2003} and references therein). We note that if the
critical magnetic field $B_c$ is finite, then the scaling behavior
occurs versus $T/(B-B_c)$ \cite{shag2017}.

\section{Violation of scaling behavior}

\begin{figure}
\begin{center}
%\vspace*{-2.5cm}
%\vspace*{-0.75cm}
\includegraphics [width=0.47\textwidth]{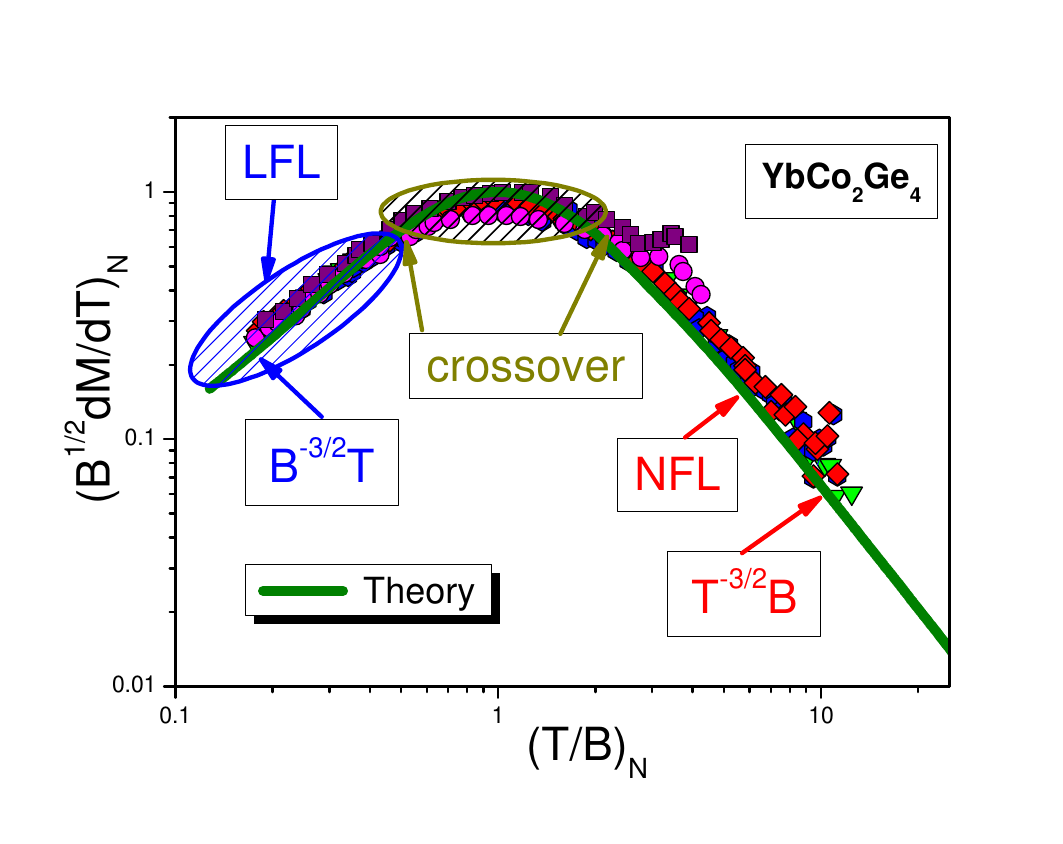}
%\vspace*{-0.75cm}
%\vspace*{2.5cm}
\end{center}
\caption{(Color online). $\rm YbCo_2Ge_4$: Scaling behavior of the
dimensionless normalized magnetization $(B^{1/2}dM(T,B)/dT)_N$
versus dimensionless $(T/B)_N$, measured at different field values
$B=0.05, 0.1, 0.2, 0.3, 0.5$ T. The data are extracted from
measurements; see Fig.~4 of Ref.~\cite{gegen:2016}. The LFL,
crossover and NFL behaviors are indicated by the arrows and hatched
areas. The theory is represented by the solid curve, describing
very well the scaling behavior of $(B^{1/2}dM(T,B)/dT)_N$ obtained
in measurements on $\rm \beta-YbAlB_4$ \cite{coleman:2015} (see
Fig.~\ref{fig5}). The LFL and NFL behaviors of
$(B^{1/2}dM(T,B)/dT)_N$ are represented by the labels $B^{-3/2}T$
and $T^{-3/2}B$, respectively.}\label{fig4}
\end{figure}

\begin{figure} [! ht]
\begin{center}
%\vspace*{-0.75cm}
\includegraphics [width=0.47\textwidth]{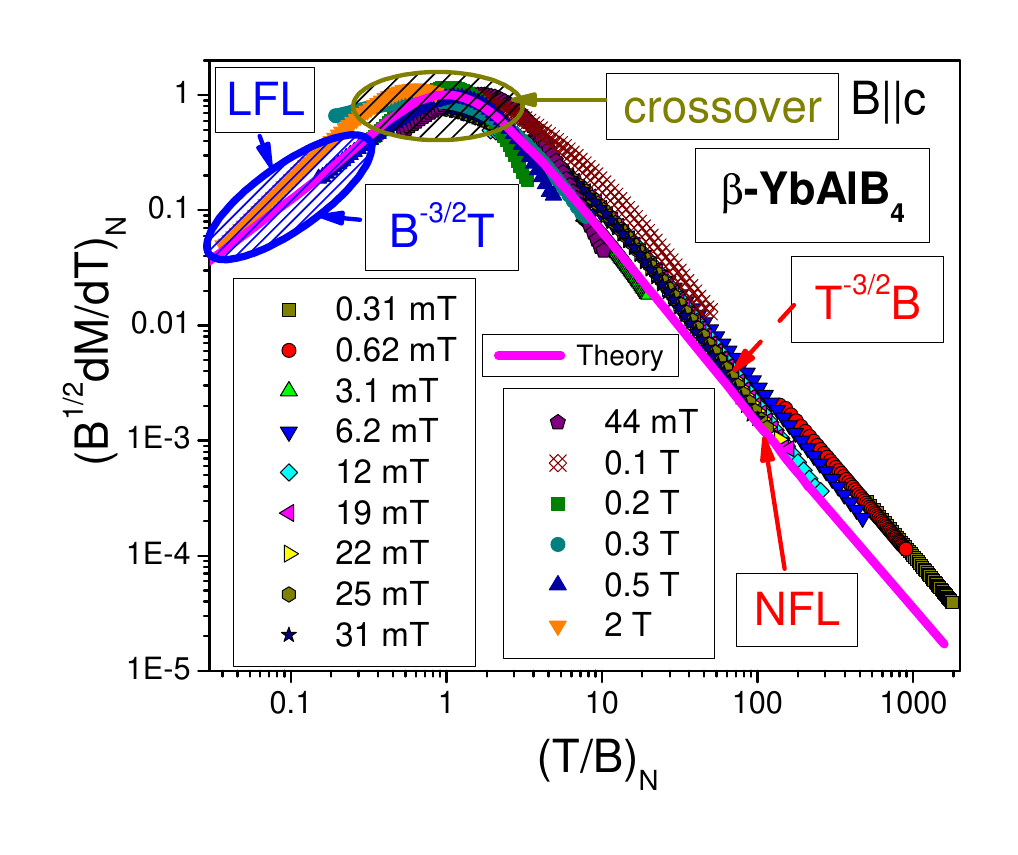}
%\vspace*{-0.75cm}
\end{center}
\caption{(color online). $\rm \beta-YbAlB_4$: Scaling behavior of
the dimensionless normalized magnetization $(B^{1/2}dM(T,B)/dT)_N$
versus the dimensionless normalized $(B/T)_N$ at different magnetic
fields ($0.31\,{\rm mT}\leq B\leq {\rm 2 \,T}$)
\cite{shaginyan:2016,bamusia:2015}. The data are extracted from
measurements on $\rm \beta-YbAlB_4$ \cite{coleman:2015}. The LFL
behavior, crossover, and NFL behavior are indicated by the arrows.
Additionally, the LFL behavior and the crossover are shown by the
hatched areas. At $(T/B)_N\gg1$ the NFL behavior is marked by the
label $T^{-3/2}B$. At $(T/B)_N\ll1$ the LFL behavior is marked by
the label $TB^{-3/2}$. The theory is represented by the solid
curve. Notation is specified in the caption of
Fig.~\ref{fig4}.}\label{fig5}
\end{figure}

Now we consider the statement that the scaling behavior can be
observed without the presence of both QCP and divergent effective
mass $M^*$ \cite{gegen:2016}. The $T/B$ scaling behaviors
experimentally observed in measurements of the magnetization
$dM/dT$ on the HF metals $\rm YbCo_2Ge_4$ and $\rm \beta-YbAlB_4$
\cite{gegen:2016,coleman:2015} are displayed in Figs.~\ref{fig4}
and \ref{fig5}. As follows from Eqs.~\eqref{UN215}, \eqref{MAXX},
and \eqref{kuka1}, the function $B^{1/2}dM/dT$ can be represented
as
\begin{equation}
\label{dMdT1}
\frac{dM}{dT}=\int^B_0\frac{\partial\chi(b,T)}{\partial T}db
\propto \int^B_0\frac{\partial M^*_N(x)}{\partial T}M^*_Mdb,
\end{equation}
where $b$ stands for magnetic field, $x=T/T_M\propto T/b$, as it
follows from Eq. \eqref{MAXX}. Again, taking into account
Eq.~\eqref{MAXX}, we obtain $M^*_M\propto b^{-1/2}$ and
Eq.~\eqref{dMdT1} then reads
\begin{equation}
\label{dMdT} B^{1/2}\frac{dM}{dT}=F(T/B),
\end{equation}
Thus, $B^{1/2}dM/dT$ is a function of the only variable $T/B$,
since $F(T/B)$ is a function of $T/B$ as it follows from Eq.
\eqref{dMdT1}.

As seen from Figs.~\ref{fig4} and \ref{fig5}, Eq.~\eqref{dMdT} and
the corresponding calculations, represented by the solid curve, are
in good agreement with the data
\cite{gegen:2016,coleman:2015,shaginyan:2016}. To calculate the LFL
behavior of $dM/dT$ taking place at $(T/B)_N\ll 1$, i.e., $T\ll B$,
we use the well-known relation $dM/dT=dS/dB$, with $S$ being the
entropy, $S\propto M^*T$. Taking into account Eq.~\eqref{MBB_15},
we obtain
\begin{equation}\label{dmdb}
\frac{dM}{dT}=\frac{dS}{dB}=T\frac{dM^*(B)}{dB}\propto -B^{-3/2}T,
\end{equation}
as shown in Figs.~\ref{fig4} and \ref{fig5}. At $(T/B)_N\gg 1$,
i.e., $B\ll T$, the system exhibits NFL behavior. Using
Eq.~\eqref{MTT_15}, we arrive at
\begin{equation}\label{dmdt}
\frac{dM}{dT}\propto\int \frac{dM^*(T)}{dT}dB\propto -T^{-3/2}B.
\end{equation}
Taking into account Eqs. \eqref{dmdb} and \eqref{dmdt}, we have at
$(T/B)_N=y\ll 1$
\begin{equation}\label{dmdx}
\sqrt{B}\frac{dM}{dT}=F(T/B)\propto y,
\end{equation}
and at $(T/B)_N=y\gg 1$
\begin{equation}\label{dmdy}
\sqrt{B}\frac{dM}{dT}=F(T/B)\propto y^{-3/2}.
\end{equation}
This theoretical result is in good agreement with experimental
observations, see Figs. \ref{fig4} and \ref{fig5}
\cite{coleman:2015,gegen:2016}. Accordingly, we conclude that the
fermion-condensation theory correctly describes the scaling
behavior, showing good agreement with the data.

\begin{figure} [! ht]
\begin{center}
%\vspace*{-1.0cm}
\includegraphics [width=0.47\textwidth]{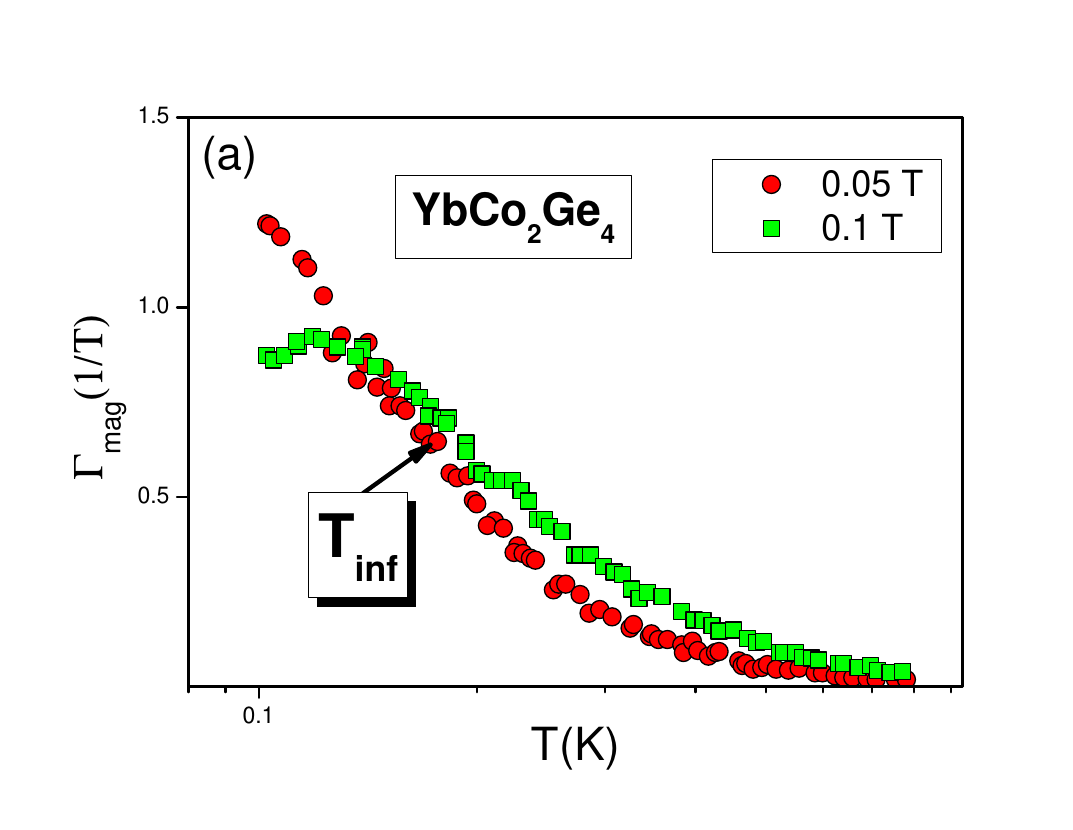}
\includegraphics [width=0.47\textwidth]{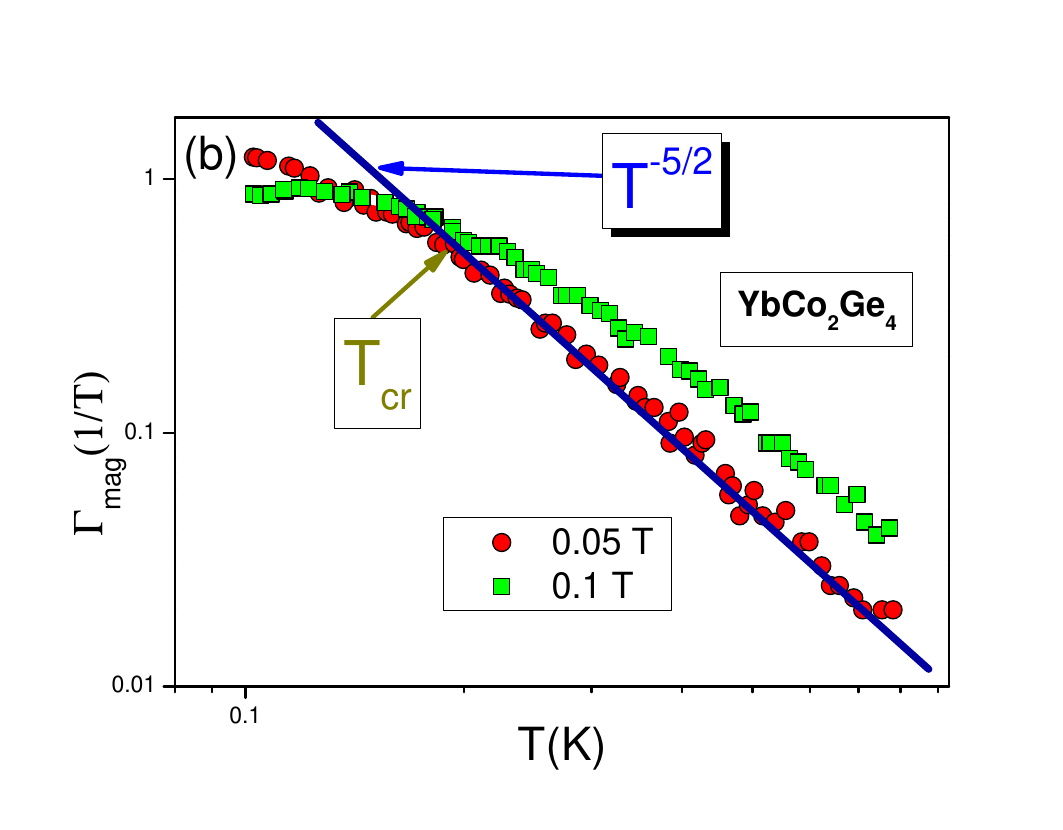}
%\vspace*{-1.0cm}
\end{center}
\caption{(color online). $\rm YbCo_2Ge_4$: Magnetic Gr\"uneisen
parameter $\Gamma_{\rm mag}(T)=-(dM/dT )/C$ versus $B$ for values
shown in the legend. The data are taken from
Ref.~\cite{gegen:2016}. (a) $\Gamma_{\rm mag}(T)$ versus a
logarithmic temperature scale. The approximate location of the
inflection point at temperature $T_{inf}$ is indicated by the
arrow. (b) $\Gamma_{\rm mag}(T)$ is shown on a double-logarithmic
plot. The solid line displays a $T^{-5/2}$ dependence at $B=0.05$
T. At $T=T_{cr}$ $\rm YbCo_2Ge_4$ enters the crossover, see Fig.
\ref{fig10}, and the dependence $T^{-5/2}$ is
vanished.}\label{fig6}
\end{figure}

Now we turn to the magnetic Gr\"uneisen parameter
$\Gamma_{mag}(T)=-(dM/dT )/C$ in the NFL regime, i.e., at $B\ll T$,
with results reported in Fig.~\ref{fig6}(a).  It is seen that
$\Gamma_{mag}(T)$ has an inflection point at $T=T_{inf}$, signaling
that there is LFL behavior at lower temperatures rather than a
divergence (see also Fig.~\ref{fig9}). Thus, we assume that the HF
metal $\rm YbCo_2Ge_4$ is located before the topological FCQPT, as
indicated by the dash-dot arrow in Fig.~\ref{fig3}. We note that
the inflection point takes place at too low temperatures and
magnetic fields, and it could not make a visible impact on the
scaling behavior reported in Fig. \ref{fig4}, that is, one needs to
carry out measurements at sufficiently low $T$ and $B$ to clarify a
possible violation of the scaling behavior. We suggest that $\rm
YbCo_2Ge_4$ can be tuned to FCQPT by the application of pressure or
by doping, as it is done in the case of $\rm CeCu_{6-x}Au_x$, while
experimental facts show that $B_c=0$ and the application of
magnetic field drives $\rm YbCo_2Ge_4$ from its QCP
\cite{gegen:2016}. The LFL behavior of $\rm YbCo_2Ge_4$ at $T\to0$
is supported by the measurements of $\Gamma_{mag}(T)$, which
exhibits divergent behavior $T^{-5/2}$ at the interval $T_{cr}\leq
T\leq 0.8$ K, as it is seen from Fig. \ref{fig6}, where $T_{cr}$ is
the crossover temperature. At $T\leq T_{cr}$ the magnetic
Gr\"uneisen parameter $\Gamma_{mag}(T)$ does not follow the
behavior indicated by the straight line because $\rm YbCo_2Ge_4$
enters the crossover region (see Figs.~\ref{fig9} and \ref{fig6}).
At the interval $T_{cr}\leq T\leq 0.8$ K one has
$\Gamma_{mag}(T)=-(dM/dT )/C\propto T^{-3/2}/T=T^{-5/2}$, since at
$T\leq 0.8$ K and $B=0$  the heat capacity $C$ demonstrates LFL
behavior, namely $C(T)\propto T$ \cite{gegen:2016}. It is seen from
Fig.~\ref{fig6} that at $T\lesssim 0.15$ K, $\rm YbCo_2Ge_4$
exhibits LFL behavior induced by the application of magnetic field
$B=0.1$. This behavior qualitatively resembles that occurring at
$B=0.05$ T.
\begin{figure} [! ht]
\begin{center}
%\vspace*{-0.75cm}
\includegraphics [width=0.47\textwidth]{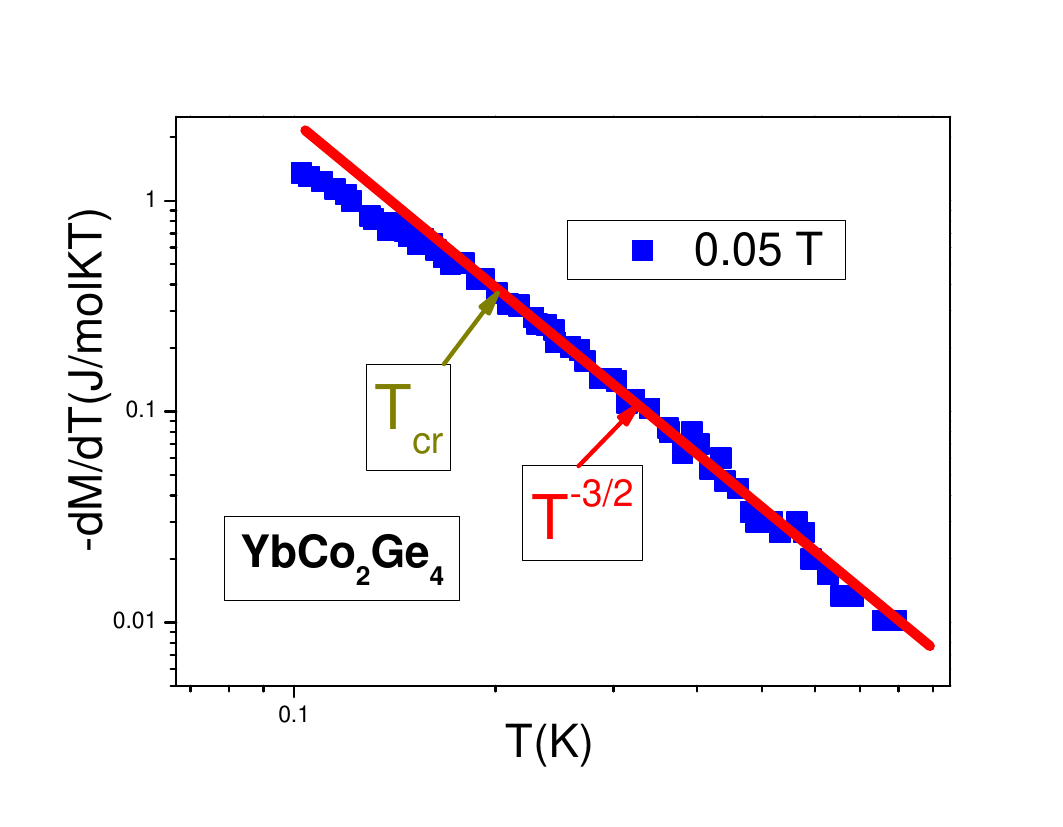}
%\vspace*{-0.75cm}
\end{center}
\caption{(color online). The $-dM/dT$ NFL behavior at fixed field
$B=0.05$ T. The solid line indicates $T^{-3/2}$ dependence on a
double-logarithmic plot. The experimental data are taken from
Ref.~\cite{gegen:2016}. The temperature $T_{cr}$ at which
the system enters the crossover region is indicated by the arrow.
See also Fig.~\ref{fig9}.}\label{fig8}
\end{figure}

Thus, we predict that at lower temperatures, $\Gamma_{mag}(T)$ will
also exhibit LFL behavior. The measurements of $dM/dT$ depicted in
Fig.~\ref{fig8} support this conclusion: at $T\geq T_{cr}$, one
finds $dM/dT\propto T^{-3/2}$ (see also Figs.~\ref{fig4} and
\ref{fig5}), while at $T\leq T_{cr}$ the divergent behavior
disappears. It is seen from Fig.~\ref{fig8} that $dM/dT$ deviates
from a straight line for $T\leq T_{cr}$, entering the crossover
region and finally exhibiting LFL behavior (see Figs.~\ref{fig9}
and \ref{fig3}). We note that the same behavior is seen in the
frustrated magnet $\rm ZnCu_3(OH)_6Cl_2$, which hosts a quantum
spin liquid and demonstrates LFL behavior at $T<400$ mK
\cite{helton}. The scaling behavior is expected to be violated in
the LFL region, whereas it would be restored with growing
temperatures $T>400$ mK \cite{shagPRB,shag:rev}. It is seen from
Fig.~\ref{fig_05_03_sc} that the scaling behavior is not violated
at $(B/T)\geq 1$, for the measurements are taken at $T\geq 1.8$ K
\cite{helton:2010}. We expect the scaling violation at $T<300$ mK
and $B<0.4$ T at the LFL behavior, see Fig.~\ref{fig_05_03_sc}. As
to $\rm YbCo_2Ge_4$, we suggest that measurements of the
thermodynamic properties at very low temperatures and magnetic
fields can clarify the physics of scaling behavior accompanied by
the divergence of the effective mass. While by now it is impossible
to exclude the possibility of the scaling behavior down to the
lowest temperatures without the $M^*$ divergence.

\section{conclusion}

The $T/B$ scaling behavior of HF compounds has been investigated at
some depth. It is shown that the HF metal $\rm YbCo_2Ge_4$ does not
exhibit scaling behavior down to lowest temperatures, since it is
located before the topological fermion condensate quantum phase
transition (FCQPT). For the same reason, the effective mass does
not diverge at the lowest temperatures. Based both on the
theoretical consideration and the experimental facts, we have shown
that there is no scaling without both the topological FCQPT and
divergence of the effective mass. We have demonstrated that HF
compounds exhibit the $T/B$ scaling down to the lowest
temperatures, provided these systems are located at the topological
FCQPT. We suggest that measurements of the thermodynamic properties
at very low temperatures and magnetic fields on $\rm YbCo_2Ge_4$
can clarify the physics of scaling behavior without the divergence
of the effective mass. We have outlined that the divergence of
effective mass $M^*$ at $T\to 0$ is of crucial importance for
projecting possible technological applications of quantum
materials. We have also demonstrated that the topological fermion
condensation theory gives a good description of the scaling
behavior of various HF compounds. As a result, the theory can be
used as well to evaluate the technological perspectives of quantum
materials. Our results are in good agreement with experimental
observations.

\section{acknowledgements}
We thank V. A. Khodel for stimulating and fruitful discussions.
This work was partly supported by U.S. Department of Energy,
Division of Chemical Sciences, Office of Basic Energy Sciences,
Office of Energy Research. J. W. Clark is indebted to the
University of Madeira for gracious hospitality during periods of
extended residence.

\end{document}